# Methods for Secondary and Tertiary Structure Prediction of Microproteins

Julio C. Facelli [1,2]

[1] Department of Biomedical Informatics and [2] Utah Clinical and Translational Science Institute, The University of Utah, Salt Lake City, Utah

**Abstract**

Microproteins are a newly recognized and rapidly growing class of small proteins, typically encoded by fewer than 100–150 codons and translated from small open reading frames (smORFs). Although research has shown that smORFs and their corresponding microproteins constitute a significant portion of the genome and proteome, there is still limited information available in the literature regarding the structural characteristics of microproteins. In this paper, we discuss the methods available for predicting their secondary and tertiary structures and provide examples of calculations done with three archetypical methods (AlphaFold, I-TASSER and ROSETTA). We present results predicting the structures of 44 microproteins. For this set of microproteins the methods considered here show a reasonable agreement among them and with the very few cases in which experimental structures are available. None-the-less, the agreement with experimental structures is not as good as for larger proteins, indicating that it is necessary to obtain a much larger set of experimental microproteins structures to better evaluate and eventually calibrate prediction methods.

**Keywords:** Protein Structure; Protein Structure Prediction, Microproteins

**Julio C. Facelli**: julio.facelli@utah.edu

**Introduction**

Microproteins are a recently identified class of proteins encoded by small open reading frames (smORFs) containing approximately 100–150 codons. Studies have shown that smORFs and their corresponding microproteins represent a substantial portion of the genome and proteome. However, there is limited information in the literature regarding the structural characteristics of these proteins. Gaining insight into their structural and functional properties is essential for understanding their biological roles. Detecting polypeptides produced by smORFs confirms their protein-coding potential by demonstrating stable protein expression and helps distinguish bona fide microproteins from small peptide fragments generated through post-translational cleavage of larger proteins (1, 2). Although the structural study of microproteins is still in its early stages, emerging research suggests that elucidating their structures and mechanisms of action could significantly enhance our understanding of their roles in various biological processes, including disease-related pathways (1-3). The state of the art in microprotein identification, function and structure has been recently reviewed (4).

Previous studies have primarily characterized microproteins by investigating their functional properties using various experimental and functional approaches (1-3). However, these methods have limitations when it comes to analyzing the structural features of microproteins, particularly in three-dimensional detail. There are very few experimental structures that have been determined for microproteins, for instance a search for the keyword "microprotein" performed Dec. 10 of 2024 in the PDB (5), show only 11 structures associated with this nomenclature and from these ones only three correspond to the isolated microproteins, therefore computational approaches are a key tool for the understanding of microprotein structures.

In this work, we explore the available methods for protein structure determination, focusing on computational structure prediction and analytical approaches. Our goal is to further elucidate the properties of microproteins as a group and to investigate whether they possess distinct structural features that differentiate them from larger proteins

Recent advances in the accuracy of protein structure prediction methods, combined with the significant experimental challenges of determining microprotein structures (6-9) highlight the importance of computational approaches in this area. Protein structure prediction techniques employ various strategies, including comparative modeling (10), de novo or ab initio methods (11), and machine learning-based approaches (12, 13) to infer three-dimensional structures from amino acid sequences.

Comparative modeling, also known as template-based modeling, relies on homologous sequences and their known crystal structures to build structural models. Ab initio prediction, on the other hand, uses empirical force fields or knowledge-based energy functions to estimate the most probable structures based on energy minimization and statistical conformational probabilities. More recently, machine learning particularly deep learning, has had a transformative impact on the field. Notably, AlphaFold (12, 13) has emerged as the leading method for protein structure prediction, as demonstrated in recent Critical Assessment of protein Structure Prediction (CASP) experiments (6, 8, 9, 14, 15).

**Methods**

Here we used three archetypical methods, I-TASSER (16-19) as exemplar for comparative modeling, ROSETTA (20-23) as exemplar for de novo/ab initio, and AlphaFold (12, 13) as exemplar for machine learning, to compare their results for a set of 44 amino acid sequences that have been reported as genuine microproteins in the previous literature (2, 24). This paper reports on the results of using these computational modeling methods to characterize the structural properties of 44 unique microproteins. The microproteins were selected following these criteria:

- The sequences contain less than 100 amino acids in length
- Proteins are single domain i.e. not a part of a larger sequence
- Proteins have been reported as microproteins in the published literature (1-3)

The FASTA files for the 44 microproteins sequences considered here were manually downloaded from UniProt (25). When possible, we compare the predicted structures with their experimental ones (or for similar amino acid sequences) that were found in the PDB as October 16$^{th}$ 2024. The microproteins studies here are listed in Table 1.

**Table 1**. Microproteins Considered in this Study. Adapted from Reference (24). Original data was manually downloaded from Uniprot (25).

| Full Name | UniProt ID | Length | Sequence |
|---|---|---|---|
| 60S ribosomal protein L41 | P62945 | 25 | MRAKWRKKRMRRLKRKRRKMRQRSK |
| Apelin receptor early endogenous ligand | P0DMC3 | 54 | MRFQQFLFAFFIFIMSLLLISGQRPVNLTMRRKLRKHNCLQRRCMPLHSRVPFP |
| ATP synthase subunit epsilon, mitochondrial | P56381 | 51 | MVAYWRQAGLSYIRYSQICAKAVRDALKTEFKANAEKTSGSNVKIVKVKKE |
| Beta-theraphotoxin-Cm1a | P84507 | 33 | DCLGWFKSCDPKNDKCCKNYTCSRRDRWCKYDL |
| B melanoma antigen 4 | Q86Y28 | 39 | MAAGAVFLALSAQLLQARLMKEESPVVSWWLEPEDGTAL |
| Small integral membrane protein 22 | K7EJ46 | 83 | MAVSTEELEATVQEVLGRLKSHQFFQSTWDTVAFIVFLTFMGTVLLLLLLVVAHCCCCSSPGPRRESPRKERPKGVDNLALEP |
| Conotoxin PIVE | P0C2C5 | 24 | DCCGVKLEMCHPCLCDNSCKNYGK |
| DDIT3 upstream open reding frame | P0DPQ6 | 34 | MLKMSGWQRQSQNQSWNLRRECSRRKCIFIHHHT |
| Dolichyl-diphosphooligosaccharide-protein glycosyltransferase subunit 4 | P0C6T2 | 37 | MITDVQLAIFANMLGVSLFLLVVLYHYVAVNNPKKQE |
| Sarcoplasmic/endoplasmic reticulum calcium ATPase regulator DWORF | P0DN84 | 35 | MAEKAGSTFSHLLVPILLLIGWIVGCIIMIYVVFS |
| Putative protein FAM66E | P0C841 | 47 | MLASGAESQDRYRNTPSSRIFTPGFRPTPATTPEPGIYMFNMEPSQP |
| Putative gastric cancer-related gene 224 protein | Q8WZA8 | 35 | MIPGNPSPGADLAVSKHFFSLSWFCGLLLLESKQK |
| Hemotin | A0A0B4K753 | 88 | MDCFKVFEVVFQSEINPLLLIPAVATIALTLCCYCYHGYQWIRDRRTARIEEQQAQLPLPLSRISITPGCSMVATTKLTHSRNSVDIY |
| Histatin-3 | P15516 | 51 | MKFFVFALILALMLSMTGADSHAKRHHGYKRKFHEKHHSHRGYRSNYLYDN |
| Humanin | Q8IVG9 | 24 | MAPRGFSCLLLLTSEIDLPVKRRA |
| Immunoglobulin lambda joining 1 | A0A0A0MT76 | 42 | PSRLLLQPSPQRADPRCWPRGFWSEPQSLCYVFGTGTKVTVL |
| Keratin-associated protein 22-1 | Q3MIV0 | 48 | MSFDNNYHGGQGYAKGGLGCSYGCGLSGYGYACYCPWCYERSWFSGCF |
| Leydig cell tumor 10 kDa protein homolog | Q9UNZ5 | 99 | MAQGQRKFQAHKPAKSKTAAAASEKNRGPRKGGRVIAPKKARVVQQQKLKKNLEVGIRKKIEHDVVMKASSSLPKKLALLKAPAKKKGAAAATSSKTPS |
| MIEF1 upstream open reading frame protein | L0R8F8 | 70 | MAPWSREAVLSLYRALLRQGRQLRYTDRDFYFASIRREFRKNQKLEDAEARERQLEKGLVFLNGKLGRII |
| MHC class I related sequence A | A0A0M3LCT1 | 23 | MGLGPVFLLLAGIFPFAPPGAAA |
| Minor histocompatibility protein HB-1 | O97980 | 41 | MEEQPECREEKRGSLHVWKSELVEVEDDVYLRHSSSLTYRL |
| Myoregulin | P0DMT0 | 46 | MTGKNWILISTTTPKSLEDEIVGRLLKILFVIFVDLISIIYVVITS |
| NLR family pyrin domain-containing protein 2B | P0DMW2 | 45 | MVSSAQLDFNLQALLGQLSQDDLCKFKSLIRTVSLGNELQKIPQT |
| Negative regulator of P-body association | A0A0U1RRE5 | 68 | MGDQPCASGRSTLPPGNAREAKPPKKRCLLAPRWDYPEGTPNGGSTTLPSAPPPASAGLKSHPPPPEK |
| Oculomedin | Q9Y5M6 | 44 | MGMYPPLLLKIYLSRHISILFYLKILYKSGIIWLSWYSFILLVL |
| Phorbol-12-myristate-13-acetate-induced protein 1 | Q13794 | 54 | MPGKKARKNAQPSPARAPAELEVECATQLRRFGDKLNFRQKLLNLISKLFCSGT |
| Cardiac phospholamban | P26678 | 52 | MEKVQYLTRSAIRRASTIEMPQQARQKLQNLFINFCLILICLLLICIIVMLL |
| Photoreceptor disk component PRCD | Q00LT1 | 54 | MCTTLFLLSTLAMLWRRRFANRVQPEPSDVDGAARGSSLDADPQSSGREKEPLK |
| Protein PIGBOS1 | A0A0B4J2F0 | 54 | MFRRLTFAQLLFATVLGIAGGVYIFQPVFEQYAKDQKELKEKMQLVQESEEKKS |
| Putative uncharacterized protein PRO2829 | Q9P1C3 | 46 | MVRPHLLKKKILGRVWWLMPVVLALWEAEVGGSLEVRSLRPAWPTW |
| Putative DNA-binding protein inhibitor ID-2B | Q14602 | 36 | MKAFSPVRSIRKNSLLDHRLGISQSKTPVDDLMSLL |

| Protein | ID | Length | Sequence |
|---|---|---|---|
| Putative glycosylation-dependent cell adhesion molecule 1 | Q8IVK1 | 47 | MKFFMVLLPASLASTSLAILDVESGLLPQLSVLLSNRLRGKTCQTGP |
| Putative makorin-5 | Q6NVV0 | 33 | MLLAAVGDDELTDSEDESDLFHEELEDFYDLDL |
| Putative tumor antigen NA88-A | P0C5K6 | 33 | MSPPSSMCSPVPLLAAASGQNRMTQGQHFLQKV |
| Sarcolipin | O00631 | 31 | MGINTRELFLNFTIVLITVILMWLLVRSYQY |
| Short transmembrane mitochondrial protein 1 | E0CX11 | 47 | MLQFLLGFTLGNVVGMYLAQNYDIPNLAKKLEEIKKDLDAKKKPPSA |
| Small cysteine and glycine repeat-containing protein 10 | A0A286YEX9 | 47 | MGCCGCGGCGGRCSGGCGGGCGGGCGGGCGGGCGGCGGGCGSYTTCR |
| Small integral membrane protein 38 | A0A286YFK9 | 51 | MTSWPGGSFGPDPLLALLVVILLARLILWSCLGTYIDYRLAQRRPQKPKQD |
| Spermatid nuclear transition protein 1 | P09430 | 55 | MSTSRKLKSHGMRRSKSRSPHKGVKRGGSKRKYRKGNLKSRKRGDDANRNYRSHL |
| Thymosin beta-15A | P0CG34 | 45 | MSDKPDLSEVEKFDRSKLKKTNTEEKNTLPSKETIQQEKECVQTS |
| | | | |
| Mitochondrial import receptor subunit TOM5 homolog | Q8N4H | 51 | MFRIEGLAPKLDPEEMKRKMREDVISSIRNFLIYVALLRVTPFILKKLDSI |
| V-alpha-13 | A0N7E1 | 23 | ASQGRKLDSYIWKRKPALLFHPY |
| Putative peptide YY-2 | Q9NRI6 | 33 | MATVLLALLVYLGALVDAYPIKPEAPGEDAFLG |

We obtained the structures predicted by the three methods using their default parameters on the University of Utah Center for High Performance Computing research clusters. In all cases we considered for the comparisons presented here the best structure as reported by the methods used. Structures for 60S ribosomal, Conotoxin, Humanin, and MHC class I related sequence A were predicted only with AlphaFold and I-TASSER because they are too short to be predicted by ROSETTA.

We used STRIDE (26) to extract the secondary structure assignments of all the structures considered here. The secondary structures were classified in an 8-state secondary structure description, including α-helix (H), residue in isolated β-bridge (B), extended strand in parallel and/or antiparallel b-sheet conformation (E), 310 helix (G), 5-helix (I), turn (T), bend (S), and coil/loops and irregular elements (C). The gyration radius of each microprotein was calculated with an R script using Bio3D (27) with each corresponding PDB file and all the comparison and structure figures were done using ChimeraX (28-30) with the default parameters.

**Results**

All FASTA files and predicted structures and their comparisons are available as .cxs files at https://zenodo.org/records/14501968. The comparison of the predicted structures using the three approaches considered here with the three available experimental ones are given in Table 2. The comparison of the predicted structures among the three methods considered here are given in Table 3 (I-TASSER and AlphaFold), Table 4 (I-TASSER and ROSETA) and Table 5 (AlphaFold and ROSETTA). Table 6 presents the comparison of the predicted structures by I-TASSER and AlphaFold, for those microproteins for which the ROSETTA structures could not be calculated. In all tables we present the values of RMS Pruned, # Pruned (number of AA in the pruned sector), % Pruned (percentage of AA in the pruned sector) and RMS All (RMS including all AA in the microprotein). The pruned values correspond to the overlap of a sequence of amino acids (larger than 3) for which the overlap is the best. Their matching quality is given by RMS Pruned and the length and relative length of the pruned segment by #Pruned, and %Pruned, respectively. Finaly the RMS All is the measure of the difference between the entire sequences of amino acids (28-30). All the RMS values are given in Å.

The analysis of the results in Table 2, shows an average RMS between predicted and experimental structures of 4.43 Å, 6.84 Å and 12.20 Å for the predicted structures using AlphaFold, I-TASSER and ROSETTA, respectively. This is in general agreement with recent CASP (15) results showing that AlphaFold is the most reliable method for protein structure prediction. But the average value of 4.43 Å is substantially larger that accuracies bellow 1 Å that have been reported for AlphaFold (12). It is difficult to make any strong conclusion of a comparison of only three structures, but the lack of microprotein structures in the AlphaFold training sets may be the root cause of this discrepancy. Clearly it is of critical importance to find a much larger set of microprotein structures to facilitate the comparison as well as retraining prediction models to better represent these systems. In other hand the results for both I-TASSER and AlphaFold for the pruned comparison show consistent sub 1 Å overlaps for pruned regions ranging from 50% to 97% of the corresponding sequences. This shows that the methods may be able to reproduce the folding of the core of the microprotein but have substantial dissimilarities at the end regions. This is apparent by examination of the pictorial comparison of structures presented in Figure 1.

Table 3 shows that comparison of the I-TASSER and AlphaFold predicted structures for the larger structures included in all the comparisons, showing good agreement, average RMS =

0.70 Å, for the pruned regions which on average are 32% of the microprotein. Similar values average RMS = 0.85 Å for the pruned region (45%) and overall average RMS = 7.8 Å are observed for the smaller microproteins (Table 6) for which ROSETTA predictions were not possible. Similar results are also observed for comparisons between I-TASSER and ROSETTA structures (Table 4), average RMS = 0.97 Å, for the pruned regions which on average are 28% of the microprotein and an overall average RMS = 11.12 Å. For the comparisons between AlphaFold and ROSETTA (Table 5), we find average RMS = 0.93 Å, for the pruned regions which on average are 26% of the microprotein and an overall average RMS = 22.60 Å. The pruned regions for which there is good overlap range from 5% to 100%.

Some examples of the structures within this range are presented in Figure 2. For Sarcolipin the percentage of pruned regions with good overlap are 100%, 65%, and 65% for the I-TASSER/AlphaFold, I-TASSER/ROSETTA and ROSETTA/AlphaFold comparisons, respectively. For B melanoma antigen 4 the percentage of pruned regions with good overlap are 51%, 26%, and 31% for the TASSER/AlphaFold, I-TASSER/ROSETTA and ROSETTA/AlphaFold comparisons, respectively. For Hemotin the percentage of pruned regions with good overlap are 32%, 22%, and 8% for the TASSER/AlphaFold, I-TASSER/ROSETTA and ROSETTA/AlphaFold comparisons, respectively. For Negative regulator of P-body association the percentage of pruned regions with good overlap are only 7%, 13%, and 9% for the TASSER/AlphaFold, I-TASSER/ROSETTA and ROSETTA/AlphaFold comparisons, respectively. All the overlaps can be found in 10.5281/zenodo.14501968, and their examination shows that, as in Figure 2 (except for Negative regulator of P-body association) there is core region with dominated by a helix structure that is reproduced reasonably well by all methods and that define the pruned region of good agreement. The end of the microprotein shows very disorganized regions and in these regions, there is much more disagreement between the methods considered here. For the Negative regulator of P-body, in which there is not helix core there is a great deal of inconsistent results for the highly disordered regions of the entire microprotein and consequently the pruned regions of good match are very small. This is consistent with the results presented previously (24) showing that the secondary structure of the microprotein are dominated by α Helix conformations.

**Table 2.** Comparison of the Predicted Structure with those Available Experimentally by NMR Measurements: Beta-theraphotoxin-Cm1a (https://www.rcsb.org/structure/6BR0#entity-1); Dolichyl-diphosphooligosaccharide-protein glycosyltransferase subunit 4 (https://www.rcsb.org/structure/2LAT#entity-1); Sarcoplasmic/endoplasmic reticulum calcium ATPase regulator DWORF (https://www.rcsb.org/structure/7MPA#entity-1). RMS values in Å.

**AlphaFold**

|  | RMS Pruned | # Pruned | % Pruned | RMS All |
|---|---|---|---|---|
| Beta-theraphotoxin-Cm1a | 0.955 | 32 | 97% | 1.103 |
| Dolichyl-diphosphooligosaccharide-protein glycosyltransferase subunit 4 | 0.997 | 18 | 49% | 7.023 |
| Sarcoplasmic/endoplasmic reticulum calcium ATPase regulator DWORF | 0.617 | 22 | 63% | 5.17 |

**I-TASSER**

|  | RMS Pruned | # Pruned | % Pruned | RMS All |
|---|---|---|---|---|
| Beta-theraphotoxin-Cm1a | 0.85 | 27 | 82% | 1.476 |
| Dolichyl-diphosphooligosaccharide-protein glycosyltransferase subunit 4 | 0.877 | 20 | 54% | 15.053 |
| Sarcoplasmic/endoplasmic reticulum calcium ATPase regulator DWORF | 0.846 | 23 | 66% | 3.995 |

**ROSETTA**

|  | RMS Pruned | # Pruned | % Pruned | RMS All |
|---|---|---|---|---|
| Beta-theraphotoxin-Cm1a | 1.369 | 18 | 55% | 4.609 |
| Dolichyl-diphosphooligosaccharide-protein glycosyltransferase subunit 4 | 0.888 | 14 | 38% | 17.432 |
| Sarcoplasmic/endoplasmic reticulum calcium ATPase regulator DWORF | 0.462 | 11 | 31% | 16.198 |

**Figure 1.** Comparison of the Predicted and Experimental Structures of **(**from Left to Right) Beta-theraphotoxin-Cm1a, Dolichyl-diphosphooligosaccharide-protein glycosyltransferase subunit 4, and Sarcoplasmic/endoplasmic reticulum calcium ATPase regulator DWORF. Pink (Experimental), Beige (AlphaFold), Blue (I-TASSER), Green (ROSETTA).

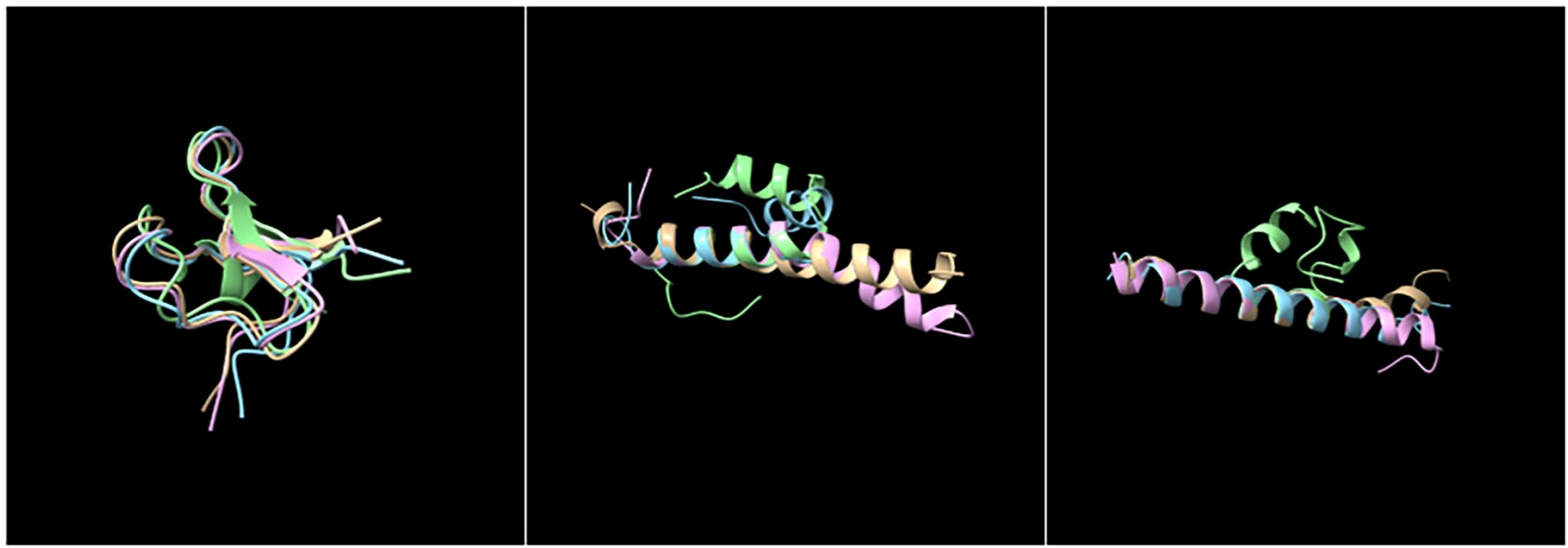

**Table 3.** Comparison of I-TASSER and AlphaFold Predicted Structures. RMS values in Å.

| | RMS Pruned | # Pruned | % Pruned | RMS All |
|---|---|---|---|---|
| Apelin receptor early endogenous ligand | 0.556 | 12 | 22% | 30.626 |
| ATP synthase subunit epsilon, mitochondrial | 0.699 | 17 | 33% | 16.215 |
| B melanoma antigen 4 | 0.448 | 20 | 51% | 25.597 |
| Small integral membrane protein 22 | 0.577 | 20 | 24% | 27.071 |
| DDIT3 upstream open reding frame | 0.373 | 22 | 65% | 12.02 |
| Putative protein FAM66E | 1.192 | 12 | 26% | 19.325 |
| Putative gastric cancer-related gene 224 protein | 0.327 | 27 | 77% | 6.556 |
| Hemotin | 0.841 | 28 | 32% | 29.722 |
| Histatin-3 | 0.794 | 15 | 29% | 41.655 |
| Immunoglobulin lambda joining 1 | 0.804 | 9 | 21% | 18.895 |
| Keratin-associated protein 22-1 | 0.911 | 5 | 10% | 36.382 |
| Leydig cell tumor 10 kDa protein homolog | 0.359 | 25 | 25% | 36.75 |
| MIEF1 upstream open reading frame protein | 0.589 | 17 | 24% | 8.702 |
| Minor histocompatibility protein HB-1 | 1.079 | 6 | 15% | 34.169 |
| Myoregulin | 0.599 | 25 | 54% | 20.54 |
| NLR family pyrin domain-containing protein 2B | 0.868 | 22 | 49% | 13.272 |
| Negative regulator of P-body association | 1.491 | 5 | 7% | 37.757 |
| Oculomedin | 0.255 | 16 | 36% | 20.903 |
| Phorbol-12-myristate-13-acetate-induced protein 1 | 0.451 | 20 | 37% | 27.711 |
| Cardiac phospholamban | 0.821 | 14 | 27% | 21.702 |
| Photoreceptor disk component PRCD | 0.324 | 18 | 33% | 29.714 |
| Protein PIGBOS1 | 0.207 | 13 | 24% | 26.465 |
| Putative uncharacterized protein PRO2829 | 0.194 | 12 | 26% | 24.011 |
| Putative DNA-binding protein inhibitor ID-2B | 1.197 | 7 | 19% | 15.845 |

| Protein | Value | Count | Percent | Score |
|---|---|---|---|---|
| Putative glycosylation-dependent cell adhesion molecule 1 | 0.766 | 14 | 30% | 15.08 |
| Putative makorin-5 | 1.074 | 15 | 45% | 7.602 |
| Putative tumor antigen NA88-A | 0.279 | 3 | 9% | 30.445 |
| Sarcolipin | 0.724 | 31 | 100% | 0.724 |
| Short transmembrane mitochondrial protein 1 | 0.457 | 17 | 36% | 17.07 |
| Small cysteine and glycine repeat-containing protein 10 | 1.169 | 5 | 11% | 15.948 |
| Small integral membrane protein 38 | 0.789 | 34 | 67% | 12.563 |
| Spermatid nuclear transition protein 1 | 1.02 | 3 | 5% | 17.903 |
| Thymosin beta-15A | 0.933 | 8 | 18% | 27.513 |
| Mitochondrial import receptor subunit TOM5 homolog | 0.618 | 21 | 41% | 21.037 |
| Putative peptide YY-2 | 0.549 | 21 | 64% | 16.784 |

**Table 4.** Comparison of I-TASSER and ROSETTA Predicted Structures. RMS values in Å.

| | RMS Pruned | # Pruned | % Pruned | RMS All |
|---|---|---|---|---|
| Apelin receptor early endogenous ligand | 0.891 | 22 | 41% | 6.441 |
| ATP synthase subunit epsilon, mitochondrial | 1.088 | 14 | 27% | 11.063 |
| B melanoma antigen 4 | 0.255 | 10 | 26% | 15.381 |
| Small integral membrane protein 22 | 1.475 | 7 | 8% | 20.044 |
| DDIT3 upstream open reding frame | 1.073 | 6 | 18% | 14.612 |
| Putative protein FAM66E | 1.118 | 11 | 23% | 14.691 |
| Putative gastric cancer-related gene 224 protein | 0.639 | 7 | 20% | 16.691 |
| Hemotin | 0.77 | 19 | 22% | 24.722 |
| Histatin-3 | 1.399 | 6 | 12% | 11.018 |
| Immunoglobulin lambda joining 1 | 1.568 | 9 | 21% | 8.271 |
| Keratin-associated protein 22-1 | 1.055 | 6 | 13% | 7.132 |
| Leydig cell tumor 10 kDa protein homolog | 0.701 | 33 | 33% | 13.896 |
| MIEF1 upstream open reading frame protein | 1.283 | 13 | 19% | 8.239 |
| Minor histocompatibility protein HB-1 | 0.908 | 7 | 17% | 8.789 |
| Myoregulin | 0.959 | 16 | 35% | 10.431 |
| NLR family pyrin domain-containing protein 2B | 1.001 | 20 | 44% | 11.015 |
| Negative regulator of P-body association | 1.534 | 9 | 13% | 11.231 |
| Oculomedin | 0.596 | 14 | 32% | 14.478 |
| Phorbol-12-myristate-13-acetate-induced protein 1 | 0.591 | 15 | 28% | 7.141 |
| Cardiac phospholamban | 1.127 | 21 | 40% | 8.233 |
| Photoreceptor disk component PRCD | 1.131 | 6 | 11% | 9.801 |
| Protein PIGBOS1 | 0.612 | 20 | 37% | 7.858 |
| Putative uncharacterized protein PRO2829 | 1.132 | 6 | 13% | 10.616 |
| Putative DNA-binding protein inhibitor ID-2B | 1.11 | 7 | 19% | 8.757 |

| Protein | | | | |
|---|---|---|---|---|
| Putative glycosylation-dependent cell adhesion molecule 1 | 0.843 | 10 | 21% | 10.404 |
| Putative makorin-5 | 0.863 | 14 | 42% | 7.24 |
| Putative tumor antigen NA88-A | 0.698 | 12 | 36% | 5.841 |
| Sarcolipin | 0.533 | 20 | 65% | 8.588 |
| Short transmembrane mitochondrial protein 1 | 0.785 | 19 | 40% | 5.023 |
| Small cysteine and glycine repeat-containing protein 10 | 1.479 | 9 | 19% | 9.075 |
| Small integral membrane protein 38 | 0.878 | 29 | 57% | 8.239 |
| Spermatid nuclear transition protein 1 | 1.261 | 5 | 9% | 20.577 |
| Thymosin beta-15A | 0.664 | 10 | 22% | 12.182 |
| Mitochondrial import receptor subunit TOM5 homolog | 1.148 | 19 | 37% | 9.002 |
| Putative peptide YY-2 | 0.712 | 14 | 42% | 12.413 |

**Table 5.** Comparison AlphaFold and ROSETTA Predicted Structures. RMS values in Å.

| | RMS Pruned | # Pruned | % Pruned | RMS All |
|---|---|---|---|---|
| Apelin receptor early endogenous ligand | 0.43 | 20 | 37% | 35.338 |
| ATP synthase subunit epsilon, mitochondrial | 1.149 | 17 | 33% | 14.596 |
| B melanoma antigen 4 | 0.484 | 12 | 31% | 27.371 |
| Small integral membrane protein 22 | 0.927 | 27 | 33% | 29.109 |
| DDIT3 upstream open reding frame | 1.211 | 4 | 12% | 17.956 |
| Putative protein FAM66E | 1.522 | 6 | 13% | 21.301 |
| Putative gastric cancer-related gene 224 protein | 0.785 | 8 | 23% | 19.639 |
| Hemotin | 1.393 | 7 | 8% | 28.39 |
| Histatin-3 | 0.823 | 6 | 12% | 34.628 |
| Immunoglobulin lambda joining 1 | 1.204 | 4 | 10% | 24.655 |
| Keratin-associated protein 22-1 | 0.665 | 5 | 10% | 34.704 |
| Leydig cell tumor 10 kDa protein homolog | 1.337 | 4 | 4% | 26.45 |
| MIEF1 upstream open reading frame protein | 1.345 | 27 | 39% | 4.279 |
| Minor histocompatibility protein HB-1 | 0.611 | 5 | 12% | 39.475 |
| Myoregulin | 0.789 | 17 | 37% | 21.483 |
| NLR family pyrin domain-containing protein 2B | 0.693 | 17 | 38% | 15.054 |
| Negative regulator of P-body association | 0.878 | 6 | 9% | 34.096 |
| Oculomedin | 0.7 | 25 | 57% | 16.263 |
| Phorbol-12-myristate-13-acetate-induced protein 1 | 0.582 | 18 | 33% | 30.424 |
| Cardiac phospholamban | 0.476 | 13 | 25% | 21.609 |
| Photoreceptor disk component PRCD | 0.824 | 20 | 37% | 25.661 |
| Protein PIGBOS1 | 1.18 | 14 | 26% | 19.766 |
| Putative uncharacterized protein PRO2829 | 1.209 | 9 | 20% | 21.486 |
| Putative DNA-binding protein inhibitor ID-2B | 1.263 | 6 | 17% | 18.485 |

| Protein | | | | |
|---|---|---|---|---|
| Putative glycosylation-dependent cell adhesion molecule 1 | 1.354 | 12 | 26% | 17.23 |
| Putative makorin-5 | 1.358 | 6 | 18% | 11.68 |
| Putative tumor antigen NA88-A | 0.556 | 4 | 12% | 30.408 |
| Sarcolipin | 0.67 | 20 | 65% | 8.316 |
| Short transmembrane mitochondrial protein 1 | 0.901 | 17 | 36% | 16.354 |
| Small cysteine and glycine repeat-containing protein 10 | 1.117 | 7 | 15% | 10.437 |
| Small integral membrane protein 38 | 0.941 | 29 | 57% | 17.583 |
| Spermatid nuclear transition protein 1 | 0.964 | 7 | 13% | 26.645 |
| Thymosin beta-15A | 0.8 | 9 | 20% | 25.957 |
| Mitochondrial import receptor subunit TOM5 homolog | 0.679 | 18 | 35% | 22.376 |
| Putative peptide YY-2 | 0.809 | 16 | 48% | 21.765 |

**Table 6.** Comparison AlphaFold and I-TASSER Predicted Structures, for which ROSETTA Structures could not be Calculated. RMS values in Å.

| | RMS Pruned | # Pruned | % Pruned | RMS All |
|---|---|---|---|---|
| 60S ribosomal protein L41 | 0.335 | 24 | 96% | 1.337 |
| Conotoxin PIVE | 1.196 | 5 | 21% | 7.263 |
| Humanin | 1.219 | 3 | 13% | 14.67 |
| MHC class I related sequence A | 0.654 | 12 | 52% | 8.023 |

**Figure 2.** Superposition of Structures within the 5% to 100% Range of the Pruned Regions with Good Overlap. From UL clockwise, Sarcolipin (100%, 65%, 65%); B melanoma antigen 4 (51%, 26%, 31%); Hemotin (32%, 22%, 8%); Negative regulator of P-body association (7%, 13%, 9%). Between parenthesis are the % of pruned regions in the comparison between I-TASSER/AlphaFold, I-TASSER/ROSETTA, and ROSETTA/AlphaFold, respectively. Beige (AlphaFold), Blue (I-TASSER), Pink (ROSETTA).

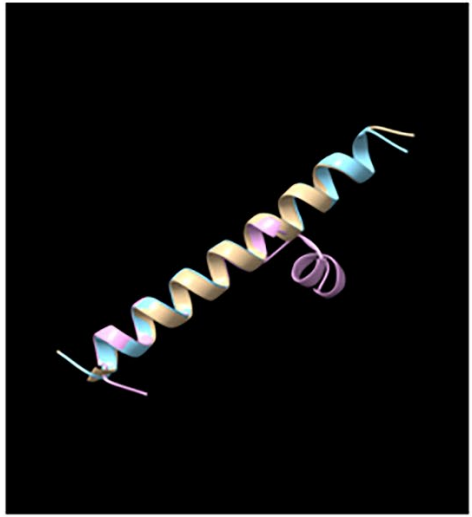
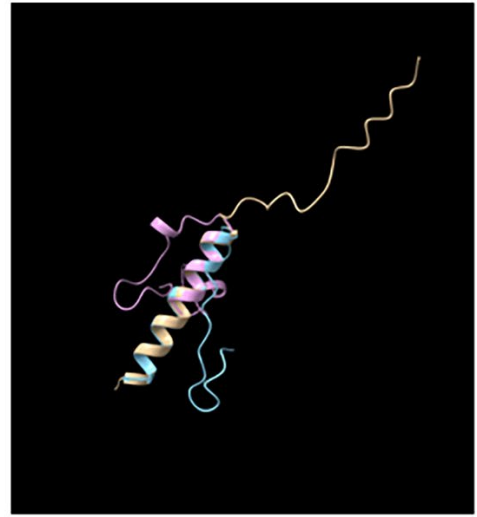
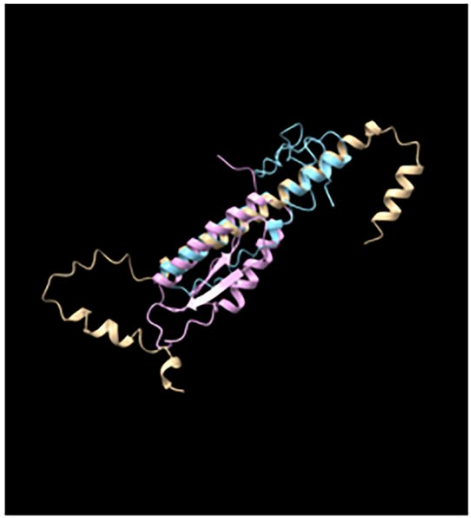
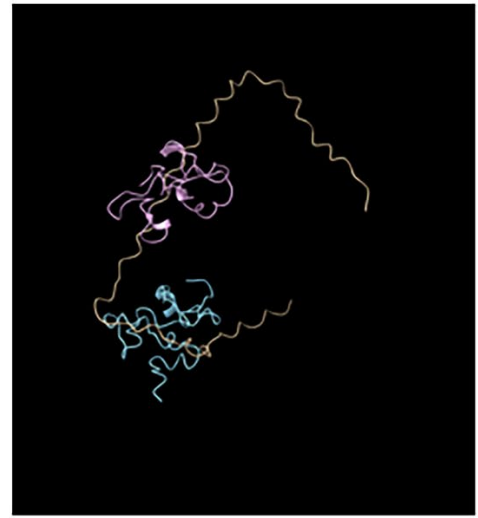

Figure 3 presents the comparison of the secondary structures predicted by the three methods considered here with the available experimental ones. The rest of the companions have been uploaded to 10.5281/zenodo.14501968. It is observed that there is a general agreement between the predicted and experimental secondary structures for the limited cases in which those are available. Except for Beta-theraphototoxin-Cm1a, the structures are dominated by α-Helix folding, which is the case (see below) for most of the other microproteins studied here. Both AlphaFold and ROSETTA predict some regions with $3_{-10}$Helix folding that are not present in the experimental ones.

The inspection of the secondary structures for the microproteins studied here for which there are no experimental counterparts are given in the repository: https://zenodo.org/records/14501968 and the following observations can be reported: there is reasonable overall agreement between the three archetypical methods used here. Overall, the structures are dominated by α-Helix folding, with some minor $3_{-10}$Helix. Both the N and C terminus of the microprotein tend to exhibit turn and coil folding, which are also found linking the predominant α-Helix segments. There are a few exceptions to these overall observations,
DDIT3 upstream open reding frame, Putative protein FAM66E, Hemotin, Humanin, Immunoglobulin lambda joining 1, Keratin-associated protein 22-1, Minor histocompatibility protein HB-1, Negative regulator of P-body association, Small cysteine and glycine repeat-containing protein 10, for which at least one of the methods used here predicts the exitance of some β sheets and in some cases large regions dominated by coils and turns.

**Figure 3.** Comparison of Secondary Structures for the Microproteins Studied here with Experimental Structures.

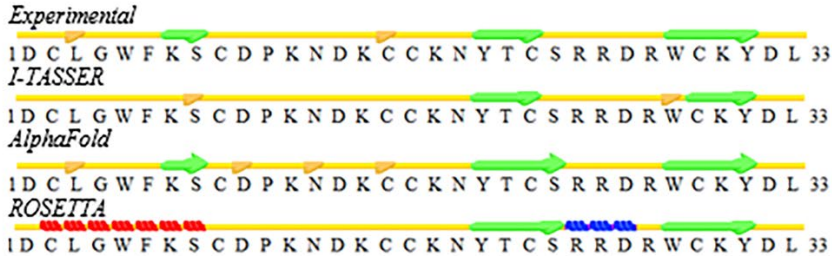

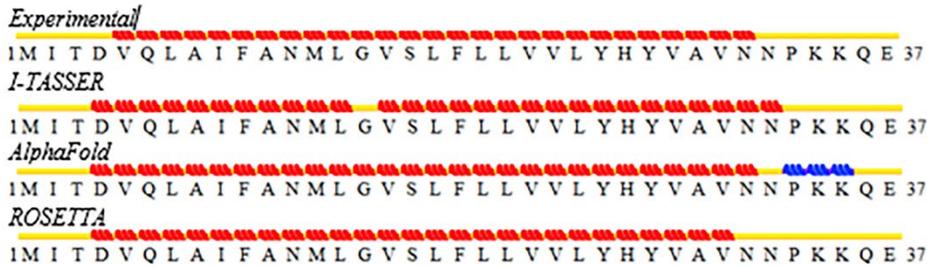

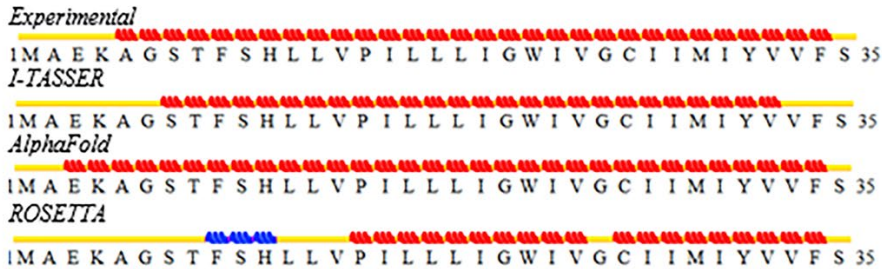

Legend of secondary structure icons:
- H Alpha-Helix
- E Extended Configuration (Beta-sheet)
- B Isolated Beta Bridge
- b Isolated Beta Bridge (Type 3 Fig 4,cd)
- T Turn
- C or " " Coil
- G 3-10 Helix
- I Pi-Helix

Table 7 presents the comparison of calculated and experimental ratios of gyration for all the microproteins studied here. Comparing ratios of gyration is very important because it allows us to compare the compactness of the structures predicted here. Our previous work using I-TASSER had shown that for these microprotein the radius of gyration follows a similar power law than those studied by before (31, 32), showing that the ratios of gyration of microproteins behave similarly to those of regular larger ones. Here we find that the ratios of gyration average $19.9 \pm 6.3$ Å, $11.4 \pm 2.4$ Å, and $10.8 \pm 1.9$ Å, for AlphaFold, I-TASSER and ROSETTA, respectively. The values of I-TASSER show that these methods predict consistently more compact structures than AlphaFold and with an average values closer to the average for the known experimental structures of $13.3 \pm 3.6$ Å. For the tree microproteins for which experimental structures are available, Beta-theraphotoxin-Cm1a, Dolichyl-diphosphooligosaccharide--protein glycosyltransferase subunit 4 and Sarcoplasmic/endoplasmic reticulum calcium ATPase regulator DWORF the predicted values by I-TASSER and ROSETTA are in much better agreement with the experimental values than those for AlphaFold.

**Table 7.** Comparison of Calculated and Experimental Ratios of Gyration.

| | I-TASSER | AlphaFold | ROSETTA | Experimental |
|---|---|---|---|---|
| 60S ribosomal protein L41 | 12.40291 | 11.63435 | n/a | |
| Apelin receptor early endogenous ligand | 11.32369 | 25.71901 | 11.02093 | |
| ATP synthase subunit epsilon, mitochondrial | 12.08668 | 16.78897 | 10.5146 | |
| Beta-theraphotoxin-Cm1a | 9.144739 | 16.44428 | 9.480134 | 9.172642 |
| B melanoma antigen 4 | 11.00577 | 19.82223 | 9.041153 | |
| Small integral membrane protein 22 | 12.42887 | 27.68764 | 15.20708 | |
| Conotoxin PIVE | 7.89883 | 10.0576 | n/a | |
| DDIT3 upstream open reding frame | 12.2987 | 15.28182 | 9.950959 | |
| Dolichyl-diphosphooligosaccharide--protein glycosyltransferase subunit 4 | 10.48763 | 9.431157 | 9.810666 | 15.58525 |
| Sarcoplasmic/endoplasmic reticulum calcium ATPase regulator DWORF | 15.51709 | 15.85249 | 8.512639 | 15.06084 |
| Putative protein FAM66E | 10.49409 | 18.23353 | 9.961591 | |
| Putative gastric cancer-related gene 224 protein | 13.60397 | 16.57206 | 9.063545 | |
| Hemotin | 15.99669 | 28.84643 | 13.9568 | |
| Histatin-3 | 10.95275 | 17.06628 | 10.96649 | |
| Humanin | 10.38356 | 11.57637 | n/a | |
| Immunoglobulin lambda joining 1 | 9.810069 | 19.68618 | 9.853099 | |
| Keratin-associated protein 22-1 | 9.452182 | 30.11635 | 9.687816 | |
| Leydig cell tumor 10 kDa protein homolog | 18.05543 | 27.2195 | 17.33295 | |
| MIEF1 upstream open reading frame protein | 12.10099 | 13.58785 | 12.66606 | |
| MHC class I related sequence A | 7.932696 | 10.30825 | n/a | |
| Minor histocompatibility protein HB-1 | 11.02868 | 34.91297 | 10.5005 | |
| Myoregulin | 11.6991 | 20.74936 | 10.12426 | |

| Protein | | | |
|---|---|---|---|
| NLR family pyrin domain-containing protein 2B | 10.23049 | 14.6146 | 9.401111 |
| Negative regulator of P-body association | 11.11929 | 30.27523 | 11.56481 |
| Oculomedin | 9.718689 | 19.79515 | 13.44144 |
| Phorbol-12-myristate-13-acetate-induced protein 1 | 10.9935 | 26.34424 | 10.78076 |
| Cardiac phospholamban | 10.74269 | 19.87108 | 12.04222 |
| Photoreceptor disk component PRCD | 11.30422 | 22.10442 | 11.14949 |
| Protein PIGBOS1 | 11.25238 | 22.88849 | 11.78394 |
| Putative uncharacterized protein PRO2829 | 10.12429 | 22.31544 | 9.959177 |
| Putative DNA-binding protein inhibitor ID-2B | 9.361572 | 17.10463 | 9.281286 |
| putative glycosylation-dependent cell adhesion molecule 1 | 9.595026 | 30.97979 | 9.706476 |
| Putative makorin-5 | 9.902365 | 12.87667 | 8.904774 |
| putative tumor antigen NA88-A | 9.380111 | 25.18957 | 8.590735 |
| Sarcolipin | 14.11977 | 14.1449 | 11.45183 |
| Short transmembrane mitochondrial protein 1 | 10.62308 | 19.0627 | 10.88649 |
| Small cysteine and glycine repeat-containing protein 10 | 8.23885 | 13.29567 | 8.774869 |
| small integral membrane protein 38 | 16.82572 | 23.30426 | 13.83696 |
| Spermatid nuclear transition protein 1 | 16.75474 | 19.63647 | 12.01156 |
| Thymosin beta-15A | 10.54257 | 23.50567 | 9.783172 |
| Mitochondrial import receptor subunit TOM5 homolog | 11.00219 | 22.31361 | 10.71685 |
| Putative peptide YY-2 | 9.887102 | 16.64908 | 8.991697 |

**Discussion and Conclusion**

In this paper, we discussed the methods available for predicting the secondary and tertiary structures of microproteins and provided examples of calculations done with three archetypical methods (AlphaFold, I-TASSER and ROSETTA) predicting the structures of 44 genuine microproteins. The results show that for this set of microproteins the methods considered here show a reasonable agreement among them and with the very few cases in which experimental structures are available. None-the-less, the agreement with experimental structures is not as good as for larger proteins, indicating that is necessary to obtain a much larger set of experimental microproteins structures to better evaluate and eventually calibrate prediction methods. For the leading protein structure prediction method, AlphaFold, the average RMS value for the microproteins studied here of 4.43 Å is substantially larger that the accuracies bellow 1 Å that have been reported for AlphaFold (12) for larger and better characterized proteins. It is difficult to make any strong conclusion of a comparison of only three structures, but the lack of microprotein structures in the AlphaFold training sets may be the root cause of these discrepancies. Clearly it is of critical importance to find a much larger set of microprotein structures to facilitate the comparison as well as retraining prediction models to better represent these systems.

One limitation of this study is the reliance on modeling tools originally developed for larger proteins, which may introduce biases in the predictions. However, due to the limited availability of experimentally determined microprotein structures, it is currently not possible to refine these tools specifically for microproteins.

15. Kryshtafovych A, Schwede T, Topf M, Fidelis K, Moult J. Critical assessment of methods of protein structure prediction (CASP)—Round XV. Proteins: Structure, Function, and Bioinformatics. 2023;91(12):1539-49. doi: https://doi.org/10.1002/prot.26617.

16. Zhang Y. I-TASSER server for protein 3D structure prediction. BMC Bioinformatics. 2008;9:40. Epub 2008/01/25. doi: 10.1186/1471-2105-9-40. PubMed PMID: 18215316; PubMed Central PMCID: 2245901.

17. Roy A, Kucukural A, Zhang Y. I-TASSER: a unified platform for automated protein structure and function prediction. Nat Protocols. 2010;5(4):725-38.

18. Lab YZ. I-TASSER FAQ, #10. Available from: https://zhanglab.ccmb.med.umich.edu/I-TASSER/FAQ.html#10.

19. Yang J, Yan R, Roy A, Xu D, Poisson J, Zhang Y. The I-TASSER Suite: protein structure and function prediction. Nature Methods. 2015;12:7-8. doi: 10.1038/nmeth.3213.

20. Rohl CA, Strauss CE, Misura KM, Baker D. Protein structure prediction using Rosetta. Methods in enzymology. 383: Elsevier; 2004. p. 66-93.

21. Leaver-Fay A, Tyka M, Lewis SM, Lange OF, Thompson J, Jacak R, et al. ROSETTA3: an object-oriented software suite for the simulation and design of macromolecules. Methods Enzymol. 2011;487:545-74. Epub 2010/12/29. doi: 10.1016/b978-0-12-381270-4.00019-6. PubMed PMID: 21187238.

22. Das R, Qian B, Raman S, Vernon R, Thompson J, Bradley P, et al. Structure prediction for CASP7 targets using extensive all-atom refinement with Rosetta@ home. Proteins: Structure, Function, and Bioinformatics. 2007;69(S8):118-28.

23. Kim DE, Chivian D, Baker D. Protein structure prediction and analysis using the Robetta server. Nucleic acids research. 2004;32(suppl_2):W526-W31.
32